\documentclass[a4paper,12pt]{article}
\usepackage{amssymb}

\input{tcilatex}
\begin{document}

\title{Incipient Separation in Shock Wave / Boundary Layer Interactions as
Induced by Sharp Fin }
\author{Hua-Shu Dou$^{1\text{,}2}$, Boo Cheong Khoo$^{2}$, and Khoon Seng Yeo%
$^{2}$ \\
$^{1}$Temasek Laboratories\\
National University of Singapore \\
Singapore 117508, SINGAPORE \\
$^{2}$Department of Mechanical Engineering\\
National University of Singapore\\
Singapore 119260, SINGAPORE }
\date{}
\maketitle

\begin{abstract}
\baselineskip 24pt

The incipient separation induced by the shock wave/ turbulent boundary layer
interaction at the sharp fin is the subject of present study. Existing
theories for the prediction of incipient separation, such as those put
forward by McCabe (1966) and Dou and Deng (1992), can have thus far only
predicting the direction of surface streamline and tend to over-predict the
incipient separation condition based on the Stanbrook's criterion. In this
paper, the incipient separation is firstly predicted with Dou and Deng
(1992)'s theory and then compared with Lu and Settles (1990)' experimental
data. The physical mechanism of the incipient separation as induced by the
shock wave/turbulent boundary layer interactions at sharp fin is explained
via the surface flow pattern analysis. Furthermore, the reason for the
observed discrepancy between the predicted and experimental incipient
separation conditions is clarified. It is found that when the wall limiting
streamlines behind the shock wave becomes\ aligning with one ray from the
virtual origin as the strength of shock wave increases, the incipient
separation line is formed at which the wall limiting streamline becomes
perpendicular to the local pressure gradient. The formation of this
incipient separation line is the beginning of the separation process. The
effects of Reynolds number and the Mach number on incipient separation are
also discussed. Finally, a correlation for the correction of the incipient
separation angle as predicted by the theory is also given.
\end{abstract}

\baselineskip 24pt


\section{NOMENCLATURE}

$\quad \,\,A$ = parameter in the \textquotedblleft triangular
model\textquotedblright

$C_{fx}$ = component of local skin friction coefficient in the main-flow
(streamwise) direction (dimensionless)

$k$= specific heat ratio (dimensionless)

$M$=Mach number (dimensionless)

$M_{\infty }$(or $M_{1}$)= incoming Mach number (dimensionless)

$p$ = static pressure (${\normalsize N/m}^{2}$)

$R=$temperature recovery factor at wall (dimensionless)

$Re_{\theta }$= Reynolds number based on momentum thickness (dimensionless).

$T=$absolute temperature (${\normalsize K}$)

$u$ = velocity component in the main flow direction in the boundary layer ($%
{\normalsize m/s}$)

$w$ = velocity component in the cross-flow direction in the boundary layer ($%
{\normalsize m/s}$)

$x$= coordinate in the streamwise direction (${\normalsize m}$)

$y$= coordinate in the direction normal to the flat plate (${\normalsize m}$)

$z$= coordinate in the direction normal to the streamwise direction ($%
{\normalsize m}$)

$\alpha $ =turning angle of the main flow, measured relative to the
direction of velocity vector at the beginning of interaction (${\normalsize %
radian}$ ${\normalsize or}$ ${\normalsize degree}$)

$\beta =$angle measured from the incoming freestream direction, centered at
the virtual origin (near the fin apex) (${\normalsize degree}$)

$\beta _{0}=$angle of shock wave (${\normalsize degree}$)

$\gamma _{w}$ = wall shear angle, i.e., angle between the direction of the
wall limiting streamline and the external streamline direction of boundary
layer (${\normalsize degree}$)

$\theta =$momentum thickness of boundary layer (${\normalsize m}$)

$\nu $ = kinematic viscosity (${\normalsize m}^{2}{\normalsize /s}$)

$\nu $(M)=Prandtl-Meyer function (${\normalsize radian}$), see Eq.(7)

$\rho $ = density (${\normalsize kg/m}^{3}$)

$\sigma =\alpha +\gamma _{w},$ turning angle of surface streamline, i.e.,
the limiting streamline direction on the wall measured from the incoming
flow direction (${\normalsize degree}$)

$\tau $ = shear stress (${\normalsize N/m}^{2}$)

\textbf{Subscripts}

$e$ = external of the boundary layer

$i$= incipient separation

$p$ = apex of \ \textquotedblleft triangular model,\textquotedblright\ Fig.5

$w$ = at walls

$x$ = component in the direction of the main flow

$z$= component in the direction of the crossflow

\section{INTRODUCTION}

The shock wave/turbulent boundary layers interaction(SW/TBLI) problem is a
very complex flow phenomenon encountered at high speed. The dynamics of the
interaction and its mechanism constitute one of the fundamental problems of
modern aerodynamics, which is very important in the design of new generation
of flying vehicles (like unconventional UAV(unmanned air vehicle)) and fluid
machinery operating at high speeds and others. Because of its importance and
implications in aeronautical engineering, shock-wave/boundary layer
interactions have been studied variously for about the past 50 years or so.
Although remarkable progress has been achieved, there still remain
observations that cannot be satisfactorily explained and physical processes
that are not quite well understood (Dolling,2001). Computational techniques
have evolved and played an increasing role in the understanding of the \
ensuring flow physics of the interactions (Knight et al., 1992; Knight et
al., 2003). However, the cost of a full Navier-Stokes calculation is still
fairly exorbitant, especially for 3D flows. In addition, issues relating to
the accuracy from a physical point of view (besides the numerical accuracy)
for purpose of quantitative comparison to experiments and the elucidation of
the complex physical flow are not always satisfactory. Therefore, some
analytical work like the simple predictive methods is still imperative for
an initial estimate of the main flow physics based on preliminary design and
then for progress to the subsequent stages leading eventually to the final
prototype. To try to do a thorough simulation at the preliminary design
stage is just too costly and most probably ineffective.

In the past over 30 years or so, there are limited works carried out on the
separation behaviour in SW/TBLI induced by a sharp fin on a flat plate as
typified by Fig. 1 (Bogdonoff, 1987; Delery and Marvin, 1986; Green, 1970;
Panaras, 1996; Settles and Dolling, 1992; Neumann and Hayes, 2002). As the
onset or occurrence of flow separation invariably changes the topology of
the flow field, one important and critical problem to resolve is how to
judge or predict the incipient separation and its underlying flow physics. A
typical topology of the surface flow pattern is shown schematically in Fig.2
(Lu, 1993; Lu et al., 1990; Settles and Dolling, 1992) for cases of
reference.

Stanbrook (1960) is probably the first to define and state that incipient
separation takes place when the wall limiting streamlines becomes aligning
with the inviscid shock wave. Since the angle of the inviscid shock wave can
be calculated by oblique shock wave theory, the prediction of incipient
separation\ then becomes a problem of predicting the direction of the wall
limiting streamline. Based on Stanbrook's criterion, McCabe (1966) proposed
a simple inviscid theory to predict the incipient separation by calculating
the deflection of the vortex tubes caused by the lateral pressure gradient
when the boundary-layer passes through the shock. For engineering purposes,
Korkegi (1973) made approximations to the McCabe's theory via utilizing the
corrections with measured test data and hence obtained a semi-empirical
formula for incipient separation: $M_{\infty }\alpha _{i}=0.3,$ for $k=1.4$
and $M_{\infty }\geq 1.60$. Subsequently, Lu (1989) took into account the
stretching of the vortex tube when the boundary-layer passes through the
shock and improved on McCabe's theory. The calculated results of Lu agree
reasonably with Korkegi's formula. In a different approach, Zheltovotov et
al.(1987) proposed a method to predict the incipient separation in
three-dimensional (3D) interactions by utilizing a semi-empirical equation
for incipient separation of two-dimensional interaction; the latter is used
as the Mach number perpendicular to the shock in three-dimensional
interaction. The results indicate that the incipient separation angle $%
\alpha _{i}$ decreases with increasing $Re_{\theta }$. Based on the 3D
compressible boundary layer theory, Dou and Deng (1992c) proposed a method
for analyzing the secondary flow within the boundary layer and predicted the
incipient separation with Stanbrook's criterion. In their analysis,
Johnston's triangle model (1960) is employed in the boundary layer and the
Prandtl-Meyer function is used in the external flow outside the boundary
layer. This analysis appears to have more fundamental physical basis than
those by McCabe (1966), Korkegi (1973), or Lu (1989). Their results also
show the trend of $\alpha _{i}$ decreasing with increasing $Re_{\theta }$
which had been previously discussed in Lu (1993) and Leung and Squire
(1995). It may be noted that Leung and Squire's (1995) experimental data
confirmed the same trend for $\alpha _{i}$ versus $Re_{\theta }$ as in Dou
and Deng's theory. Furthermore, McCabe (1966) and Dou and Deng (1992c)\
found that the incipient separation angle $\alpha _{i}$ via skin-friction
line (also called \textquotedblleft limiting streamlines\textquotedblright\
or \textquotedblleft surface streamlines\textquotedblright\ (Lighthill,
1963)) calculation and both works correctly predicted the wall limiting
streamline direction before separation (see also Deng et al., 1994).
However, there is an overprediction of the incipient separation condition,
and Korkegi's criterion was therefore re-interpreted by Lu and Settles
(1990) to mean that significant separation occurs for $\alpha >\alpha _{i}$.
This overpredicting issue has remained unsolved or not addressed
satisfactory (Settles and Dolling, 1992). This provides the motivation of
the present work to systematically clarify the said overprediction and the
validity of reinterpretation.

It should be mentioned that the mechanism for incipient separation is still
not well understood. In the past twenty years or so, although computational
fluid dynamics (CFD) technique has made a significant progress in the
understanding of the flow physics and the structure of the flow field
(Knight et al., 1992; Thivet et al., 2001), simple analytical work is still
relevant to this subject, especially in the light of behaviour trend with
varying flow parameters. Therein lies further motivation for the present
work. In this paper, we firstly review Dou and Deng's (1992c) theory, and
compare the theoretical prediction results with the experimental data of Lu
and Settles (1990). The effect of Reynolds number is then discussed. Then,
the variation of the surface flow pattern with increasing deflection angle
is analyzed, and the mechanism of the incipient separation induced by
SW/TBLI at sharp fin is explained. It is shown that the genesis of incipient
separation is traced to by the secondary flow due to lateral pressure
gradient. In this way, it is hoped that the above mentioned discrepancies
between theories and experiments are clarified. In addition, a correlation
for the correction of $\alpha _{i}$ by the theoretical prediction is also
given.

\section{THEORY}

Dou and Deng (1992c)'s theory is based on the \textquotedblleft
skewed\textquotedblright\ boundary layer concept. We shall briefly review
this theory. In a swept shock wave/boundary layer interaction, a pressure
gradient generated by the shock wave is exerted on the boundary-layer. The
pressure gradient in three-dimensional interactions can be decomposed into
two components, i.e., the streamwise direction component and the lateral
direction component. The former generates a viscous interaction with the
incoming boundary layer, similar to two-dimensional interactions, causing
flow retardation; the latter makes the velocity profiles of the
boundary-layer skewed, and generates a secondary flow in the boundary layer,
perpendicular to the streamwise direction( see Fig. 3). The relative
magnitude of these two components indicates which of the above two phenomena
is more important in dominating the separation in 3D interactions. For the
interactions induced by a sharp fin in supersonic flow (with $M_{\infty
}\geq 2.0$), the role of the lateral component of the pressure gradient
overwhelms that of the streamwise component of the pressure gradient as
explained by Dou and Deng (1992b) (see also Dou, 1991). For high Mach
number, the incipient separation is dominated by the secondary flow because
of the small shock wave angle and the large lateral pressure gradient.

The boundary layer on the wall is skewed owing to the lateral pressure
gradient. The fluid particles near the wall travel along the path with
larger curvature than that of the inviscid flow, as shown in Fig. 3.
Starting from the outer edge of the boundary layer, the velocity vector in
the boundary layer gradually deviates from the mainflow direction, and the
deviation angle reaches a maximum at the wall. The direction of wall
limiting streamline deviates from the primary streamline by an angle $\gamma
_{w}$ (see Fig. 3). The angle $\gamma _{w}$ increases with the flow from the
the front towards the downstream direction due to the lateral pressure
gradient. When this angle reaches a certain value, the wall limiting
streamlines converge to a single line from the upstream, which can be
detected by oil film technique in experiments and is commonly called
\textquotedblleft incipient separation line\textquotedblright (Fig.2). At
this incipient separation line, the wall limiting streamline becomes
perpendicular to the direction of the local pressure gradient. This type of
separation can be broadly described by the model of Maskell (1955) and
Lighthill (1963). Generally, it is difficult to define when this
\textquotedblleft incipient separation line\textquotedblright\ appears in
the calculation, and the Stanbrook's criterion is normally just invoked for
the prediction. Since the direction of shock wave angle can be easily
calculated by inviscid shock wave equations, the prediction of incipient
separation line can be strictly done by calculating the direction of wall
limiting streamline (Fig. 4).

For this type of pressure-driven three-dimensional turbulent boundary-layer,
there is a fair amount of works which have been carried out (see White,
1974; Olcmen and Simpson, 1992). If the wall shear angle $\gamma _{w}$ is
not very large and the lateral flow is not bi-directional, Johnston's
triangular model seems to give the best approximation (White, 1974). This
said model has been widely used in many engineering problems (Smith, 1972;
Swafford and Whitfield,1985). The law of wall in the three-dimensional
turbulent boundary-layer based on this model by Johnston is still one of the
best description up to recent years(Olcmen and Simpson, 1992).

Johnston (1960) studied the nature of the boundary layer on a flat wall
under the influence of a turning main flow. It is stated by Johnston that
this type of boundary-layer problems are commonly described as
secondary-flow problems because the three-dimensional perturbations in the
layer are caused by the associated pressure gradient due to the curvature of
the main flow. The effect as generally observed is a skewing of the
boundary-layer velocity vectors toward the center of curvature of the main
flow. The type of the boundary layer in SW/TBLI is of the same nature as
that described by Johnston. Thus, the theory developed by Johnston can be
used to analyze the boundary-layer in SW/TBLI. Indeed, Lowrie(see Green,
1970) and Myring (1977) have used this model in their studies of SW/TBLI.

Johnston divided the turbulent boundary layer into two regions along the
direction of boundary layer thickness. He assumed that a collateral region
near the wall exists and the direction of the velocity vector in this region
is coincident with the shear stress vector. This region is called the inner
region of the boundary layer. In the outer region, the behavior of flow is
primarily dominated by the outer inviscid flow. According to this model, the
crossflow velocity profile of the boundary layer can be expressed as
follows(see Fig.5): 
\begin{equation}
\frac{w}{u_{e}}=\frac{u}{u_{e}}\tan \gamma _{w}\qquad \quad \quad 
{\normalsize for}\qquad \frac{u}{u_{e}}\leq \left( \frac{u}{u_{e}}\right)
_{p}\text{,}
\end{equation}%
\begin{equation}
\frac{w}{u_{e}}=A\left( 1-\frac{u}{u_{e}}\right) \qquad {\normalsize for}%
\qquad \frac{u}{u_{e}}\geq \left( \frac{u}{u_{e}}\right) _{p}\text{,}
\end{equation}%
where $\gamma _{w}$ is the angle between the wall limiting streamline and
the external streamline, $\left( \frac{u}{u_{e}}\right) _{p}$ is the
streamwise velocity ratio at the apex of the triangle. If the variation of
the direction of external flow is known, the direction of the wall limiting
streamline can be calculated by evaluating the angle $\gamma _{w}$.

From Eq.(1) and Fig.5, the following expression is obtained, 
\begin{equation}
\tan \gamma _{w}=A\left[ \left( \frac{u}{u_{e}}\right) _{p}^{-1}-1\right] 
\text{.}
\end{equation}

It can be seen that the parameters $A$ and $\left( \frac u{u_e}\right) _p$
must be determined for the calculation of $\gamma _w$.

\subsection{The Parameter \textit{A}}

Johnston (1960) expressed the parameter \textit{A} as a function of the
parameters of the main flow. For the cases in which a pressure gradient
exists along the main streamline direction($\partial p/\partial x\neq 0$)
and the turning angle varies in the normal direction of mainflow($\partial
\alpha /\partial z\neq 0$), he derived the following equation,

\begin{equation}
A=2u_{e}^{2}\int_{0}^{\alpha }\frac{d\alpha }{u_{e}^{2}}\text{,}
\end{equation}%
where $u_{e}$ is the velocity at the outer edge of the boundary-layer, $%
\alpha $ is the turning angle of main flow and is measured relative to the
main flow direction at the beginning where the streamline of main flow first
turns. The minus sign in the above equation which appears in the original
equation \ of Johnston (1960) has been neglected. This is because we only
consider the flow deflection in a single direction and are only concerned
with the magnitude of $A$.

According to the conservation of energy, the velocity ratio across the shock
wave is expressed as follows,

\begin{equation}
\frac{u_{e2}}{u_{e1}}=\frac{M_{2}}{M_{1}}\sqrt{\frac{1+\left[ \left(
k-1\right) /2\right] M_{1}^{2}}{1+\left[ \left( k-1\right) /2\right]
M_{2}^{2}}}\text{,}
\end{equation}%
where the subscripts $1$ and $2$ represent the parameter before and after
the shock wave, respectively.

At the incipient separation condition, the strength of the shock wave is
generally weak. Thus, the relation of the Mach numbers across the shock can
be approximated by the Prandtl-Meyer relation. In other words, the shock
wave at the edge of the boundary layer may be represented by a series of
isentropic processes. Using this relation, the following closed form
solution for the parameter $A$ can be easily obtained. Our calculations
showed that the error due to this approximation is very small compared to
the exact relation of shock wave. Introducing Eq.(5) into Eq.(4) and using
the Prandtl-Meyer relation, we have, 
\begin{eqnarray}
A &=&\frac{2M_{2}^{2}}{1+\left( k-1\right) M_{2}^{2}/2}\int_{0}^{\alpha }%
\frac{1+(k-1)M_{2}^{2}/2}{M_{2}^{2}}d\alpha  \nonumber \\
&=&\frac{-2M_{2}^{2}}{1+\left( k-1\right) M_{2}^{2}/2}\int_{M_{1}}^{M_{2}}%
\frac{\sqrt{M_{2}^{2}-1}}{M_{2}^{3}}dM_{2}  \nonumber \\
&=&\frac{M_{2}^{2}}{1+\left( k-1\right) M_{2}^{2}/2}\left( \text{arc cos}%
\frac{1}{M_{1}}-\text{arc cos}\frac{1}{M_{2}}-\frac{\sqrt{M_{1}^{2}-1}}{%
M_{1}^{2}}+\frac{\sqrt{M_{2}^{2}-1}}{M_{2}^{2}}\right) ,
\end{eqnarray}%
and Prandtl-Meyer relation is expressed as 
\begin{equation}
\nu \left( M\right) =\sqrt{\frac{k+1}{k-1}}\text{arc\ tan}\sqrt{\frac{k-1}{%
k+1}\left( M^{2}-1\right) }-\text{arc\ tan}\sqrt{M^{2}-1}\text{.}
\end{equation}%
The deflection angle $\alpha $ is related to Prandtl-Meyer equation by 
\begin{equation}
\alpha =\nu \left( M_{1}\right) -\nu \left( M_{2}\right) \text{ .}
\end{equation}

\subsection{The Parameter $\left( u/u_e\right) _p$}

For compressible flows, Smith (1972) made an extension of the
\textquotedblleft velocity triangle\textquotedblright\ of the
three-dimensional boundary layers for incompressible flows, which was first
given by Johnston (1960). The parameter $\left( u/u_{e}\right) _{p}$, in
Eqs.(1) and (2), was expressed as in reference (Smith, 1972),

\begin{equation}
\left( \frac{u}{u_{e}}\right) _{p}=\overline{y}_{p}\sqrt{\frac{\rho _{e}}{%
\rho _{w}}\frac{C_{fx}\text{cos}\gamma _{w}}{2}}\text{,}
\end{equation}%
where $\overline{y}_{p}=\frac{y_{p}}{\nu _{w}}\sqrt{\frac{\tau _{w}}{\rho
_{w}}}$ and $C_{fx}$ is the component of the skin friction coefficient in
the direction of main flow. The density ratio $\rho _{e}/\rho _{w}$ can be
calculated from the energy equation for compressible flow. The components of
the wall shear stresses are depicted in Fig.3.\ Besides Smith (1972), \ this
equation has been used by Myring (1977) and Swafford and Whitfield (1985).

From the physical relationship for the three-dimensional boundary-layer, the
value of $\overline{y}_{p}$ is related to the conditions of boundary layer.
In Johnston (1960) and Smith (1972), $\overline{y}_{p}=14.0$ is employed
based on the test data obtained at low-speed. For the case of high-speed
boundary layers (supersonic flows), the velocity distribution across the
layer is very different from that of low speed flows. The role of viscosity
is confined to within the thinner layer next to the wall. The test results
at $M_{1}=3.0$ by Settles was introduced in reference (Delery, 1985). The
experiments showed that both the relative height of the viscous-layer and
the relative height of sonic line decrease with the increasing Reynolds
number, and the latter drops faster than the former. Therefore, at high
Reynolds number (high Mach number), the velocity vector at the edge of the
viscous-layer has larger deflection. For the turbulent boundary-layer at
supersonic flow range ($Re$ is very high), the value of $\overline{y}_{p}$
is expected to be less than $14$ ($\overline{y}_{p}<14$). According to the
structure and the velocity distribution of turbulent boundary-layer, it is
assumed in this study that $\overline{y}_{p}$ is respondent to the
intersection point of the viscous-sublayer and the layer of logarithmic law
defined in the two-dimensional turbulent boundary-layer, i.e. $\overline{y}%
_{p}=11$ (Kuethe and Chow, 1986).

For the selction of $\overline{y}_{p}=11,$ a detailed expalnation is
provided as follows. As discussed above, within the boundary layer, it can
be divided into two regions in the direction normal to the wall: region I
which is near the wall and region II which adjacent to the external flow. In
region I, the layer is very thin and the flow deflection is constant along
the direction normal to the wall. Thus, the velocity vector is along the
same direction in this region. This region is generally called
\textquotedblleft collateral flow.\textquotedblright\ In region II, the
layer is skewed and the flow direction varies gradually. The velocity vector
changes from the direction of main (streamwise) flow at the edge of boundary
layer to the direction of wall streamline in region I.

In region II, the role of viscosity is very small and the behaviour is
almost controlled by the external flow. The degree of skewness of the flow
in this layer can be described by the equation, $\partial p/\partial r=\rho
u^{2}$/$r$. Since the pressure is constant in the direction normal to the
wall within the layer, the value of $\partial p/\partial r$ is constant too.
Thus, the variation of velocity will lead to the variation of radius of
curvature of the streamline. As such, the fluid with lower velocity will
have smaller radius of curvature of the associated streamline. Thus, the
flow becomes more deflected as it approaches the wall, and the layer is
skewed. In region I, the flow is mainly controlled by viscosity and is
almost not influenced or affected by the main flow. The viscous stress is
larger in this layer and this gives rise to the variation of pressure
distribution. In this layer, the governing equation normal to the streamline
can be expressed as $\partial p/\partial r=\rho u^{2}$/$r+\partial \tau
_{r\theta }/(r\partial \theta )$. The viscous stress balances a part of the
centrifugal force and thus the pressure gradient is reduced approaching
towards the wall. Therefore, this much wider variation of pressure could
lead to a reduction of deflection within this layer. As a result, the flow
direction in this layer experiences almost no change and a \textquotedblleft
collateral flow\textquotedblright\ is formed within this viscous layer.
Since the flow in the viscous sub-layer (also called linear layer when
expressed in terms of the inner layer variable) is viscous dominated, it is
not unusual to expect that the \textquotedblleft collateral
flow\textquotedblright\ extends to the whole height of the sub-layer, i.e., $%
\overline{y}_{p}=11$. Therefore, it is deemed reasonable to assume that $%
\overline{y}_{p}=11$, at the junction point of two layers for high speed
flows. 

Substituting Eq.(9) into Eq.(3), the following equation can be derived, 
\begin{equation}
\tan \gamma _{w}=A\left( \frac{0.13}{\sqrt{\frac{\rho _{e}}{\rho _{w}}%
C_{fx}\ \text{cos}\gamma _{w}}}-1\right) \text{.}
\end{equation}

The density ratio is obtained from the energy equation,

\begin{equation}
\frac{\rho _{e}}{\rho _{w}}=\frac{T_{w}}{T_{e}}=1+\frac{k-1}{2}RM_{2}^{2}%
\text{,}
\end{equation}%
where $R$ is the temperature recovery factor at the wall. For the adiabatic
turbulent boundary layer, it takes on $0.88$ generally. For $2\leq M_{1}\leq
6$, the following correlation gives a better approximation to the
experimental data (Dou and Deng, 1992c), 
\begin{equation}
R=0.80+0.01\left( 8.0-\frac{M}{2}\right) \text{.}
\end{equation}

\subsection{The Coefficient of Skin Friction}

The local skin friction coefficient towards the streamline direction was
taken from that for the 2D flow and given as (Dou, 1991),

\begin{equation}
\frac{C_{fx}}{C_{fxi}}=\left( 1+0.13M^{2}\right) ^{-0.73}\text{,}
\end{equation}%
where $C_{fxi}$ is the coefficient of skin friction for incompressible
turbulent boundary layer on flat plate. In this study, the
Karman-Schoenherr's equation recommended by Hopkins and Inouye (1971) is
utilized. This equation is applicable to the whole range of the Reynolds
number of turbulent boundary layer and given as,

\begin{equation}
\frac{1}{C_{fxi}}=17.08\left( \text{log}_{10}Re_{\theta }\right) ^{2}+25.11%
\text{log}_{10}Re_{\theta }+6.012\text{ .}
\end{equation}

\subsection{Calculation of Shock Wave Angle}

The shock wave angle can be calculated by the implicit oblique shock wave
theory (e.g., Kuethe and Chow, 1986), or by the following approximate
equation given by Dou and Deng (1992a). This latter formula is of
satisfactory accuracy over a wide ranges of the upstream Mach number and
flow deflection angle. For $\alpha \leq 15^{o}$ and $2\leq M_{1}\leq 5$, the
relative error to the exact value is less than 1\% and given as,

\begin{equation}
\tan \beta _{0}=\frac{1}{\sqrt{M_{1}^{2}-1}}+\frac{k+1}{4}\frac{M_{1}^{4}}{%
\left( M_{1}^{2}-1\right) ^{2}}\alpha +\frac{1}{2}\left( \frac{k+1}{4}%
\right) ^{2}\frac{M_{1}^{6}\left( M_{1}^{2}+4\right) }{\left(
M_{1}^{2}-1\right) ^{3.5}}\alpha ^{2}\text{.}
\end{equation}

Using Eqs.(6) to (15), the variation of $\gamma _{w}$ along the streamwise
direction can be calculated for a given Mach number and $Re_{\theta }$ for
the incoming flow with increasing $\alpha $. Next, the turning angle $\sigma 
$ of the surface streamline on the wall due to the action of shock
disturbance can be evaluated. When the turning angle $\sigma $ at the wall
equals to the shock angle $\beta _{0}$, the separation of the
three-dimensional boundary layer is considered to have occurred as was shown
by Stanbrook (1960). Similar calculations can be carried out for various
incoming flow conditions.

In shock wave/boundary layer interactions, since the streamlines converge
from the upstream, the boundary layer becomes thickening along the
streamwise direction within the interaction region. Thus, the streamlines
are also considerably deflected away from the wall. The actual streamline
deflection is not just lying in a plane parallel with the surface as assumed
in the analysis. However, because the normal velocity to the wall in the
boundary layer is generally very much smaller compared to the streamwise
velocity, the wall-normal deflection has been neglected in this analysis
owing to its small effect. The reason for this approximation can be
explained as follows in detail. In the boundary layer theory, this normal
velocity is usually at least one order of magnitude smaller than the
streamwise component. In present study, the shock wave generated by the high
speed flow (fluid flow above the sonic line) acts on the boundary layer. The
shock wave in the inviscid flow is a plane with a jump in flow parameters.
When this plane interacts on the boundary layer, the section of the
interaction zone starting from the leading edge of the interaction to the
incipient separation line is very short. This is because the boundary layer
in high speed flows is usually very thin. Within this short distance, the
variation of normal velocity is also not large; note that the flow is not
completely separated from the surface. Therefore, the influence of the
normal velocity can be neglected in the analysis. 

\section{RESULTS AND DISCUSSION}

\subsection{Comparison of the Theories with Experiments}

The experimental data on incipient separation\ were generally obtained by
oil film visualization technique in tunnel experiments (Deng et al., 1994;
Dou and Deng, 1992b; Dou and Deng, 1992c; Lu and Settles, 1990; Settles and
Dolling, 1992). Detailed description of the experiments can be found in
these works. Generally, the test model of a sharp fin is amounted on one
flat plate and placed in the supersonic wind tunnel (Fig.1). Before the
experiment, a layer of oil film is spread thinly on the flat plate. When the
air flow passes the plate, the oil film moves from the upstream to the
downstream, and oil streaks are formed along the streamlines on the flat
plate. After the wind tunnel is shut down, one can obtain this oil streak
pattern by taking photograph or by using transparency glue papers. These
pictures have the features as presented schematically in Fig.2. The
occurrence of incipient separation was mostly decided in terms of the
formation of the convergent line of the wall limiting streamlines from the
upstream as according to Lighthill's criterion (Lighthill, 1963). Figure 6
shows the comparison of the experimental data reported by Lu and Settles
(1990) and the predictions using Dou and Deng's method for four Mach
numbers. The intersection point of the turning angle $\sigma $ of surface
streamlines with the shock wave angle $\beta _{0\text{ }}$corresponds to the
condition set by Stanbrook's criterion (A-A line), i.e., the wall limiting
streamlines becomes parallel to the inviscid shock wave. The agreement of $%
\sigma (\equiv \alpha +\gamma _{w})$ value between the theory (Eq.(6) to
(10)) and the experiments is very good before the incipient separation
occurrence based on Stanbrook's criterion. The arrows at the abscissa
indicate the incipient separation as judged or ascertained using Korkegi's
equation (B-B line); this was reported by Lu and Settles (1990). The latter
(B-B line) is a little lower than those obtained using Stanbrook's criterion
(A-A line). On the other hand, the incipient separation angles reported in
experiments are even lower than Korkegi's value (B-B line) (see Korkegi,
1973). Besides this, it should be mentioned that Deng et al (1994) found
that McCabe's theory concurs fairly well with the experimental data of Lu
and Settles (1990) for the turning angle of wall limiting streamlines, but
for the incipient separation. The above mentioned inconsistencies between
theory and experiments have left much to be desired. Furthermore, the
physical mechanism for the occurrence/initiation of the incipient separation
is still yet to be fully understood. It is the intent of this work to
provide a reasonable explanation for such observation and also to present a
necessary and yet robust correction to achieve overall consistency.

Figure 7 shows the comparison of the results predicted by Dou and Deng
(1992c)'s theory as well as others (see also Dou and Deng, 1992c). In this
figure, the effect of $Re_{\theta }$ on $\alpha _{i}$ is also displayed.
When $Re_{\theta }$ is increased, $\alpha _{i}$ becomes smaller. This
implies that the boundary layer is more easily susceptible to separation at
higher $Re_{\theta }$. This is somewhat similar to that found for
two-dimensional interactions (Delery, 1985). In two-dimensional shock
wave/boundary layer interactions (Delery, 1985; Delery and Marvin, 1986),
most experiments showed that the resistance to separation (increasing
strength of shock wave) decreases with the $Re_{\theta }$ number at low to
moderate values of $Re_{\theta }.$ Then, there is a small reversal of the
incipient separation condition at high $Re_{\theta }$ number. For
three-dimensional shock wave/boundary layer interactions, the correlation
results for two-dimensional interactions indicated the following effect of
Reynolds number $Re_{\theta }$: the incipient separation angle $\alpha _{i}$
decreases with increasing $Re_{\theta }$ (Zheltovodov et al., 1987). Leung
and Squire (1995) have discussed the $Re_{\theta }$ influence on incipient
separation, and their experimental data confirmed the same tendency of $%
\alpha _{i}$ as that predicted with Dou and Deng's theory. However, McCabe's
theory does not depict this influence of Reynolds number (It should be
pointed out that the Reynolds number may vary when the Mach number changes
in real flows). Almost all of the data in Fig.7 were obtained for incoming
flow Reynolds number of about $10^{4}$ ($Re_{\theta }=5\times 10^{3}\symbol{%
126}5\times 10^{4}$). It can be seen that the incipient separation angles
from the experiments are lower than those predicted by McCabe's and Dou and
Deng's theory.

\subsection{Analysis of Surface Flow Patterns}

The process of the formation of the incipient separation could be described
by analyzing the evolution of the surface flow patterns. A typical schematic
of surface flow from flow visualization experiments is shown in Fig. 2. The
surface flow pattern formed by a sharp fin is conical (Lu, 1993; Lu et al.,
1990; Settles and Dolling, 1992; Oudheusden et al., 1996), and not
cylindrical (Johnston, 1960; Myring, 1977; Knight et al., 1992; Van
Oudheusden et al., 1996). The topological pattern of the surface streamlines
in conical interactions has been interpreted recently by Van Oudheusden et
al. (1996)$.$ In Fig.2, the schematic surface pattern depicts a geometric
conicity, in that rays are shown emanating from a certain origin close to
the fin apex. The direction of the oil streaklines are to a large extent
fairly independent of the distance along the ray. In this kind of conical
flow field, all the flow quantities take on constant \ values on the ray
through the conical center of the flow field. The variation of surface flow
pattern with the increase of the strength of shock wave for a given incoming
Mach number could be divided into the following six stages as shown in the
schematic diagram of Fig. 8. This figure is summarized from a large quantity
of experimental data on surface flow patterns in the literature (McCabe,
1966; Korkegi, 1973; Kubota and Stollery, 1982; Zheltovodov et al, 1987; Lu
and Settles, 1990; Settles and Dolling, 1992; Lu, 1993; Deng and Liao, 1993;
Deng et al., 1994; Leung and Squire, 1995). In Fig.8, for simplicity, it is
assumed that the virtual origin of the conical region coincides with the
apex of the fin on the plate. The evolution of the surface streamlines with
the increasing strength of shock wave is shown up to the formation of the
primary separation line. The following observations are made:

(a) The deflection angle is small and the shock wave is weak, and the effect
of secondary flow is negligible.

(b) On increasing the deflection angle, the wall limiting streamlines behind
the shock turn to the shock wave trace gradually, but the shock wave
strength is not large enough to deflect the surface streamline to make it
parallel to the shock. The main feature of this stage is the gathering of
surface streamlines from the upstream behind the shock wave.

(c) Further increasing the deflection angle, the wall limiting streamlines
converge and coalesce onto a single line from the upstream. This is true
when the strength of the shock wave is still not large enough to deflect the
surface streamlines behind the shock wave to become parallel to the shock
wave. This single line formed from the upstream is just the
\textquotedblleft \emph{incipient separation line}\textquotedblright\
exhibited by oil streak pattern technique in experiments, which symbolizes
the beginning of the separation process.

(d) Upon further increasing the deflection angle, the \textquotedblleft
incipient separation line\textquotedblright\ formed from the upstream
rotates (shifts) continuously with the increasing shock wave angle.
Meanwhile, the surface streamlines behind the shock wave becomes parallel to
the shock wave. This is the condition defined by the Stanbrook's criterion.
Of course, this condition arrives later than the appearance of
\textquotedblleft incipient separation line\textquotedblright\ indicated by
experiments (Fig.8c). This is the reason why the theories overpredict the
occurrence of the incipient separation compared with the experiments shown
in Fig. 6 and Fig. 7.

(e) On further increase of the deflection angle, the \textquotedblleft
incipient separation line\textquotedblright\ formed from the upstream
rotates (shifts) continuously with the increasing shock wave angle, and the
wall limiting streamlines from the downstream of the shock wave also
converge to this line from another side.

(f) When the deflection angle is increased to a certain value, the
\textquotedblleft incipient separation line\textquotedblright\ also becomes
one\ convergent line of the wall limiting streamlines from the downstream;
this is called the \textquotedblleft primary separation
line.\textquotedblright\ Some authors prefer to use the appearance of
primary separation line to judge the separation (Kubota and Stollery, 1982).

Now, the whole process of the formation of convergent line in the oil film
pattern can be described as follows. In the experiments in a supersonic
tunnel (Lu and Settles, 1990), the color oil substance is normally painted
on the test plate, on which the sharp fin model is amounted. When the air
flow passes the plate, the film with the oil substance is moved from the
upstream to the downstream location by the air flow. However, the oil film
moving from upstream will be blocked by this ray of incipient separation
line. This is because the surface streamlines forms a half-closed pattern
ahead of this ray. Thus, the oil film behind this ray will be eventually
dispersed. The oil film ahead of this ray will still be kept on the plate.
This is the reason why the incipient separation line can be formed in their
experiments. Any two streamlines from the upstream and the convergent line
form a \textquotedblleft U\textquotedblright\ -shaped like pattern. As such,
this half-closed loop existing towards the downstream could prevent the oil
film flow from passing through to the convergent line.

\subsection{Physical Mechanism for Incipient Separation}

The three-dimensional separations induced by SW/TBLI at the sharp fin can be
described by the model of Maskell (1955) or Lighthill (1963). According to
Lighthill's criterion, it is considered that the boundary layer is separated
when the wall limiting streamlines converge to a single line. In terms of
mass conservation, the converging from the upstream to a single line is
enough to be considered as flow separation as also discussed by Kubota and
Stollery (1982). From the point of view of the equilibrium of forces, the
vector of skin-friction force is perpendicular to the direction of local
pressure gradient at the incipient separation line. Therefore, \emph{when} 
\emph{the skin-friction lines of incoming flow becomes perpendicular to the
direction of the local pressure gradient}, the formation of the
\textquotedblleft incipient separation line\textquotedblright\ becomes
possible. In fact, Stanbrook's criterion satisfies this condition. However,
with the gradual increase of the shock wave angle, the flow state as
expressed by Stanbrook's criterion is not the first manifestation of this
condition. This condition is satisfied indeed somewhat earlier, as is
observed in for case(c) of Fig.8. This is strictly the main mechanism for
the formation of incipient separation line.

Although the surface features of the cylindrical and conical interactions
are different (Setteles and Dolling, 1992), they do share some common
properties/features. For both the cylindrical and conical interactions to be
possible, it is required the direction of the skin-friction line at
incipient separation line be perpendicular to the local pressure gradient.
However, for the conical interaction, it is not necessary for the
\textquotedblleft incipient separation line\textquotedblright\ to align with
the shock wave, while it is so for the cylindrical interaction. The \emph{%
interaction region} on the flat plate generated by SW/TBLI at a sharp fin is
a conical zone, which stretches across the inviscid shock wave trace on the
plate when the shock wave is weak (Fig.8a to Fig.8c). Thus, the maximum of
the turning of the surface streamlines as well as the primary convergence
line is behind the shock wave, and this convergence line makes an angle to
the inviscid shock line (Fig.8c). As a result, this convergence line is
formed before it becomes parallel to the shock as the shock wave angle
increases. Van Oudheusden et al. (1996) argued that the far field of the
conical interaction does not possess a quasi-two dimensional structure in
the cross-flow plane of the radial direction. They showed that the conicity
of the inviscid flow regions in supersonic flow produces a geometrically
conical surface flow pattern. There is essential difference between the
cylindrical and conical interactions.

From the above discussions we can say that \emph{when} \emph{the wall
limiting streamlines (not only one streamline) behind the shock wave becomes
aligning with one ray from the virtual origin (near the fin apex) as the
strength of shock wave increases, the incipient separation line is generated.%
} At this ray, the direction of the skin-friction vector is perpendicular to
the local pressure gradient. The wall limiting streamlines of incoming flow
then converge and coalesce to this ray. Thus, this ray could prevent the oil
film from spreading across it and therefore can be easily detected in
experiments with oil streak pattern technique. In most experiments, this ray
is considered as the incipient separation line as associated with
Lighthill's criterion.

If the flow is cylindrical as analyzed in Inger (1986) and Myring (1977),
there exists none of the stage between case(c) and case (d) in Fig.8 for
conical interactions; in other words, case (d) coincides with case (c). The
incipient separation line formed from the upstream uniquely corresponds to
that defined by Stanbrook's criterion (Inger, 1986; Myring, 1977). However,
for conical interactions, there is an angle difference between case(c) and
case (d) in Fig.8. Therefore, the process for the formation of the primary
separation line for conical interactions is very different from that for
cylindrical interactions. As a results, the Stanbrook's criterion is
applicable to cylindrical interactions, but is not directly applicable to
conical interactions. This is why there is still a discrepancy between the
prediction using this criterion and the experimental data.

\subsection{Correction for Incipient Separation Angle}

The difference of the deflection angle between the cases(c) and (d), $\Delta
\alpha ,$ is shown in Fig.9 for two set of data. It can be found that $%
\Delta \alpha $ decreases with the increasing Mach number. This is in accord
with the physical mechanism of the interaction because the pressure ratio
across the shock wave increases with the Mach number, and the pressure ratio
at separation line is almost constant for three-dimensional separation of
supersonic flows (Dou, 1991; Dou and Deng, 1992b).

Since it is difficult to find a criterion to define the condition of
Fig.8(c), we still take Stanbrook's criterion as the incipient separation
criterion, and add a correction to the predicted incipient separation angle $%
\alpha _{i}.$

Assume that the correction $\Delta \alpha $ is only related to the Mach
number. Thus, we can work out a correlation using the experimental data for
the correction to the theoretical prediction. The result is shown below and
in Fig.9, 
\begin{equation}
\Delta \alpha =0.20M_{1}^{2}-1.8M_{1}+5.70\qquad \text{for }1.6<M_{1}<5\text{%
.}
\end{equation}%
The corrected incipient separation angle $\alpha _{ic}$ is, 
\begin{equation}
\alpha _{ic}=\alpha _{i}-\Delta \alpha \qquad \text{for }1.6<M_{1}<5\text{.}
\end{equation}

\section{CONCLUSIONS}

The conclusion can be summarized as follows:

1. Dou and Deng (1992c) developed a theoretical method to analyze the
three-dimensional turbulent boundary-layer in SW/TBLI. Based on this theory,
we carried out the predictions for the experiments of Lu and Settles
(1990).\ Dou and Deng's theory is physically founded to be better than those
previous works of McCabe (1966), Korkegi (1973) or Lu (1989) in allowing a
fuller understanding of the secondary flow influence and the effect of $%
Re_{\theta }$. The prediction of the wall limiting streamline direction by
this theory yielded good agreement with Lu and Settles' (1990) experimental
data.

2. The \textquotedblleft incipient separation line\textquotedblright\ formed
from the upstream is generated by the secondary flow induced by the lateral
pressure gradient. The incipient separation line provided by the experiments
is the condition of the first appearance of the wall limiting streamline
perpendicular to the local pressure gradient. \emph{When the wall limiting
streamlines from the upstream becomes aligning along with one ray from the
virtual origin (near the fin apex) as the strength of shock wave increases,
the incipient separation line is formed, at which the wall limiting
streamline is perpendicular to the local pressure gradient.}

3. The process for the formation of the primary separation line for conical
interactions is very different from those for cylindrical interactions. The
Stanbrook's criterion is only applicable to cylindrical interactions, and
not applicable directly to conical interactions. The disagreement between
the prediction by this criterion and the experiments is attributed to the
intrinsic behavior of conical interaction.

4. The difference of the deflection angle for incipient separation between
the prediction and the experiments, $\Delta \alpha ,$ decreases with the
increasing Mach number. A correlation equation for the correction to the
theoretical predicted incipient separation angle $\alpha _{i}$ is
given.

\bigskip

\end{document}